\documentclass[a4paper,11pt]{article}
\usepackage{pos}
\usepackage{lineno}

\title{Quarkonia production and elliptic flow in small systems measured with ALICE}

\manuallySeparateAuthors
\author*[a]{Maurice Coquet} 
\author{on behalf of the ALICE collaboration}

\affiliation[a]{Université Paris-Saclay Centre d’Etudes de Saclay (CEA), IRFU, Département de
Physique Nucléaire (DPhN), Saclay, France}

\emailAdd{maurice.louis.coquet@cern.ch}

\abstract{The production of quarkonia in hadronic collisions provides a unique testing ground for understanding quantum chromodynamics (QCD) since it involves both the perturbative and non-perturbative regimes of this theory. Given that a satisfactory description of quarkonia production has not yet been achieved, new measurements that can  provide new insights, helping to constrain models, are needed. The ALICE apparatus allows to measure inclusive J/$\psi$ production, as well as to separate prompt charmonia from those originating from b-hadron decays. The study of the azimuthal correlation of the emitted particles, e.g. via the measurement of the elliptic flow ($v_2$), in high multiplicity proton-proton (pp) collisions can probe collective behaviour in small systems.
In this contribution, we present new measurements of the inclusive, prompt and non-prompt J/$\psi$ production in pp collisions at different collision energies, together with the J/$\psi$ $v_2$ in high multiplicity pp collisions at $\sqrt{s}$=13 TeV.}

\FullConference{%
  41st International Conference on High Energy physics - ICHEP2022\\
  6-13 July, 2022\\
  Bologna, Italy
}

\begin{document}
\maketitle

\section{Introduction}

Quarkonium production involves a combination of soft and hard processes which probe different regimes of Quantum Chromodynamics (QCD): the production of heavy quarks, which occurs in initial hard-scattering processes for which the large quark mass sets a hard scale, can be described by perturbative QCD (pQCD) calculations, whereas the binding of the quark-antiquark pairs into colourless final states is a non-perturbative processe. The separation and treatment of these scales differ in available quarkonium-production models (see Ref.~\cite{Lansberg_2020} for an overview). Another important aspect of quarkonia studies in small systems is the search for collectivity in high-multiplicity events. Indeed, striking similarities between small systems (high multiplicity pp or p--Pb collisions) and Pb--Pb collisions have been seen for observables usually considered as quark-gluon plasma (QGP) signatures, such as strangeness enhancement~\cite{2017}, or long-range azimuthal correlations, which indicate the presence of collective phenomena~\cite{Abelev_2013}. In this regard, the study of flow observables in the heavy-flavour sector, in particular of the J/$\psi$ elliptic flow, is important to further establish and characterize flow in small systems. Finally, the measurement of quarkonium production in small systems also serves as a reference to study heavy-ion collisions and the QGP.

With the ALICE apparatus, quarkonium can be measured either at midrapidity $(|y|<0.9)$ in the dielectron decay channel, or at forward rapidity $(2.5<y<4)$ in the dimuon decay channel. The main detectors used at midrapidity for the measurements presented in these proceedings are the Inner Tracking System (ITS), used for tracking, vertexing, as well as for the measurement of midrapidity charged-particle multiplicity, and the Time Projection Chamber (TPC) used for tracking and particle identification via the measurement of the specific energy loss. At forward rapidity, the measurements are performed with the muon spectrometer, composed of absorbers, a dipole magnet, tracking and triggering stations, which provide reconstruction and identification of muon tracks. A minimum bias trigger, beam-gas background rejection, and measurement of multiplicity at forward rapidity, are assured by the V0 detector, which spans the pseudorapidity ranges $-3.7<\eta<-1.7$ (V0C) and $2.8<\eta<5.1$ (V0A). Although inclusive quarkonium at both forward and midrapidity can be measured down to zero $p_{\mathrm{T}}$, only the detectors at midrapidity offer sufficient vertexing performance to separate prompt and non-prompt charmonium. A detailed description of the ALICE apparatus can be found in Ref.~\cite{ALICE:2008ngc}. 

\section{Results}

\begin{figure}[h]
\begin{center}
\rotatebox{0}{
\includegraphics[width=0.5\linewidth]{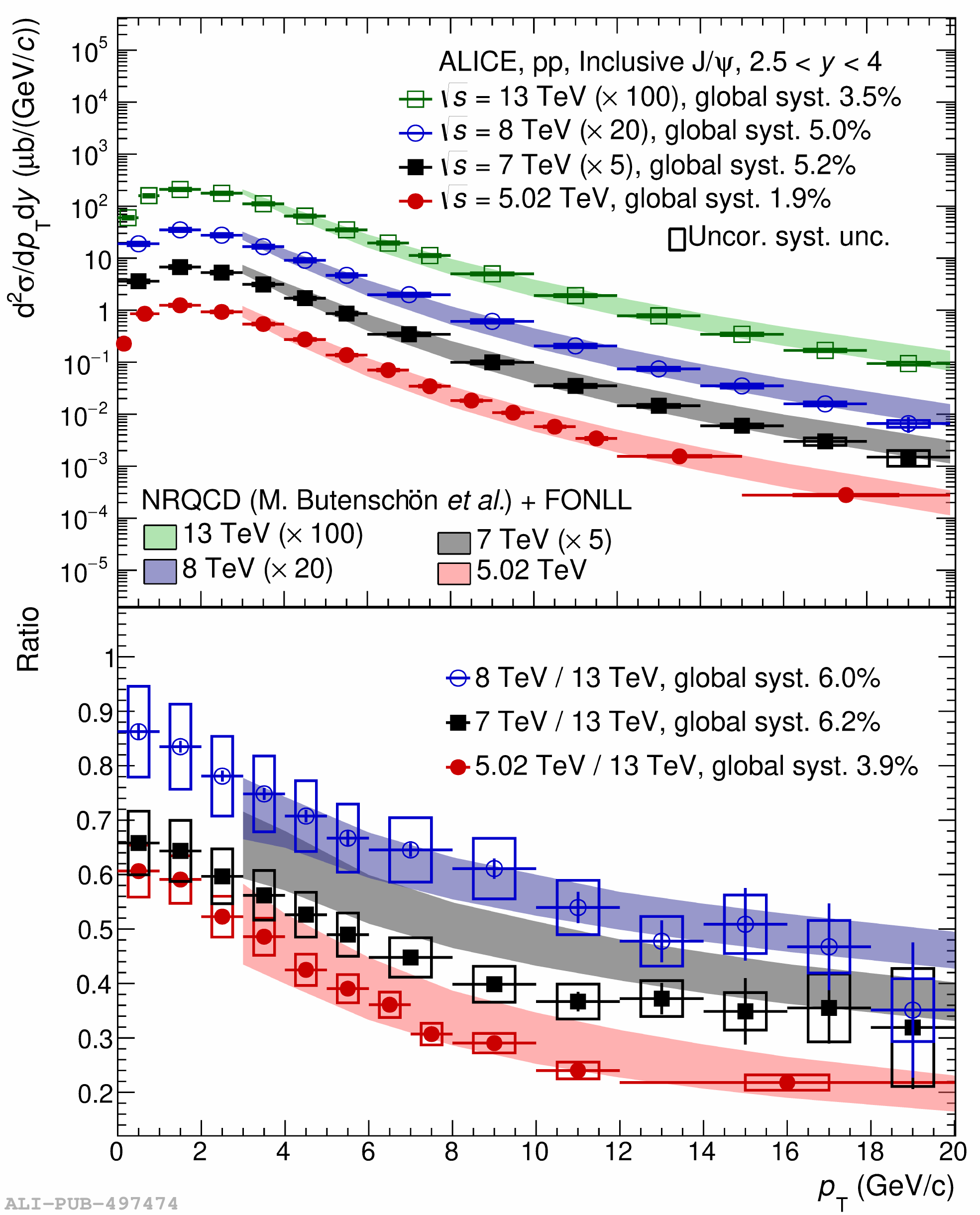} 
\includegraphics[width=0.5\linewidth]{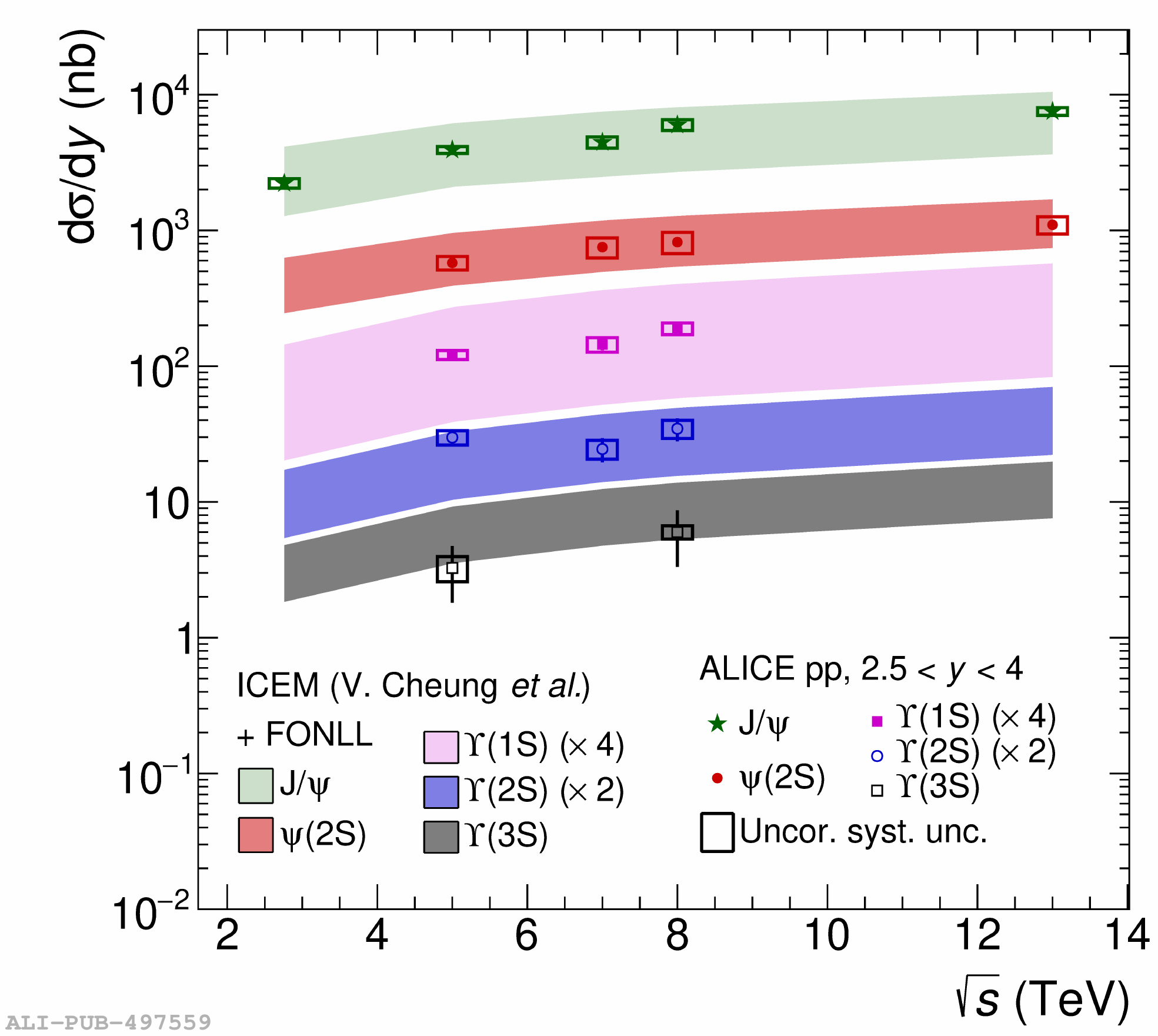} }
\end{center}
\caption{Left: (top) inclusive J/$\psi$ $p_{\mathrm{T}}$-differential production cross section at forward rapidity at different collision energies  and (bottom) ratios of J/$\psi$ cross sections at different energies to the $\sqrt{s}=13$ TeV cross section~\cite{inclquark}, compared to NRQCD+FONLL calculation~\cite{Butensch_n_2011, Cacciari_1998}. Right: J/$\psi$, $\psi$(2S), $\Upsilon$(1S), $\Upsilon$(2S), and $\Upsilon$(3S) production cross sections per unit of rapidity as a function of the collision energy, compared with theoretical calculations from ICEM+FONLL~\cite{Cheung_2018, Cacciari_1998}. }
\label{fig:incljpsi}
\end{figure}

The inclusive $\mathrm{J} / \psi$ $p_{\mathrm{T}}$-differential cross sections at forward rapidity measured by ALICE, at collision energies $\sqrt{s}=5.02$, 7, 8 and 13 TeV~\cite{inclquark} are presented in Fig.~\ref{fig:incljpsi} (top left). The measurement at $\sqrt{s_{\mathrm{NN}}} = 5.02$ TeV uses a data sample larger by about a factor of ten with respect to that used for the earlier measurement~\cite{Acharya_2017}. An increase of the $\mathrm{J} / \psi$ cross section is observed with collision energy. In addition, the data is compared to the expectation of a two-component model obtained by summing a Non-Relativistic Quantum Chromodynamics (NRQCD) calculation~\cite{Butensch_n_2011} describing the prompt component of the cross section, and a Fixed-Order-Next-to-Leading-Logarithm (FONLL) calculation~\cite{Cacciari_1998} which accounts for the non-prompt contribution from b-hadron feed-down. 
Within uncertainties, the measured cross sections are in good agreement with these models for $p_{\mathrm{T}} \geq 3$ GeV/$c$ at all considered collision energies. Figure~\ref{fig:incljpsi} (bottom left) shows the ratios of the inclusive $\mathrm{J} / \psi$ cross sections at different collision energies to the $\sqrt{s} = 13$ TeV one. In these ratios the theoretical and experimental uncertainties partly cancel, leading to stronger constraints on models. A good agreement within uncertainties between data and NRQCD+FONLL is observed, except for the 7-to-13 TeV ratio which is slightly overestimated by the model. In addition, a hardening of the inclusive $\mathrm{J} / \psi$ $p_{\mathrm{T}}$-differential spectrum is observed at $\sqrt{s}=13$ TeV with respect to lower energies. This hardening can be interpreted as the combination of two factors: an increase of the prompt $\mathrm{J} / \psi$ mean $p_{\mathrm{T}}$ with collision energy, and an increase of the non-prompt $\mathrm{J} / \psi$ fraction with collision energy, as predicted by FONLL. 

Figure~\ref{fig:incljpsi} (right) shows the inclusive production cross sections of $\mathrm{J} / \psi$, $\psi\mathrm{(2S)}$ and of the three $\Upsilon$ states measured at forward rapidity as a function of the collision energy~\cite{inclquark}. The data are reproduced within uncertainties by ICEM~\cite{Cheung_2018} + FONLL~\cite{Cacciari_1998} calculations, over the whole inspected collision-energy interval.

\begin{figure}[h]
\begin{center}
\rotatebox{0}{
\includegraphics[width=0.52\linewidth]{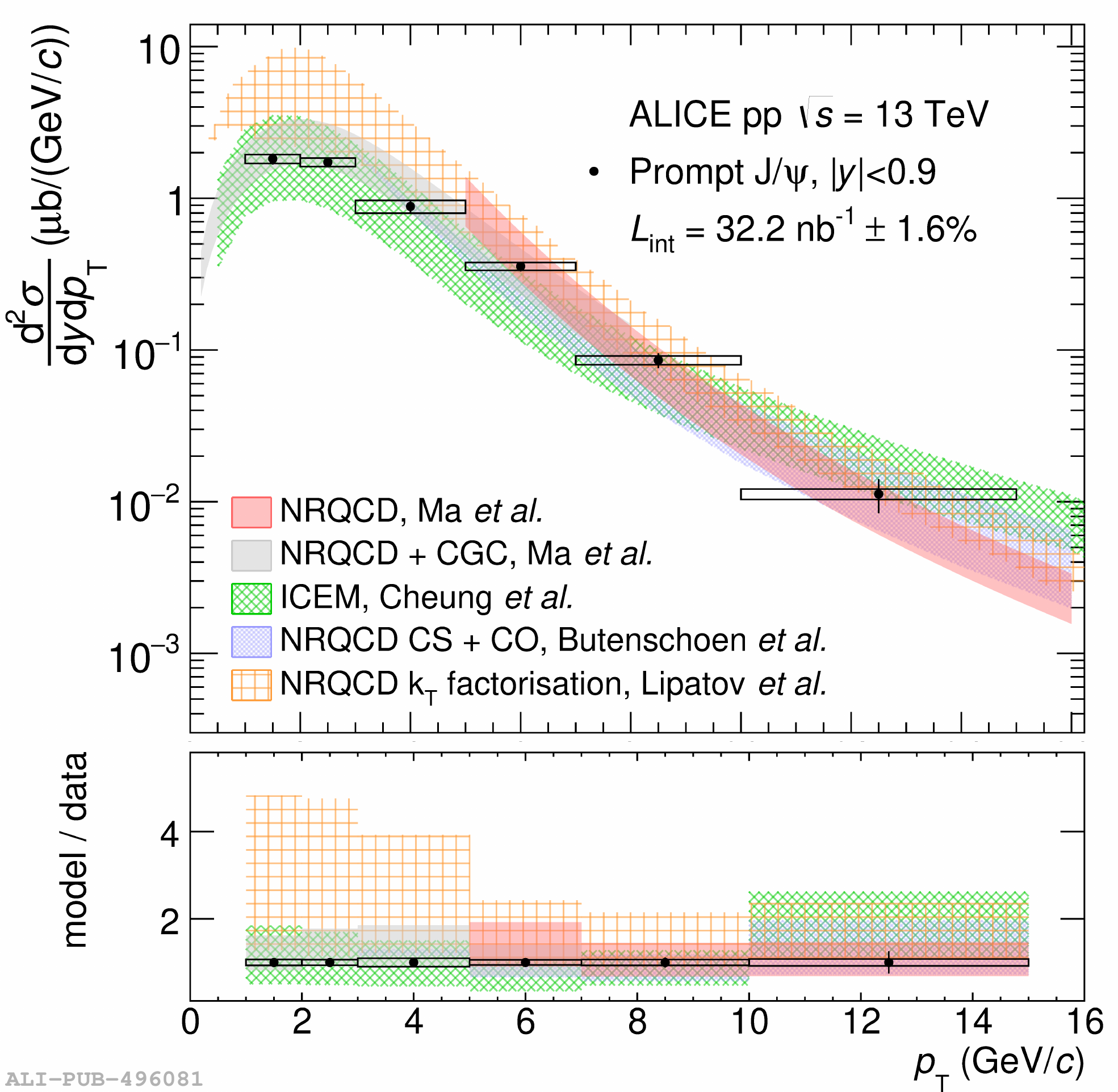} 
\includegraphics[width=0.52\linewidth]{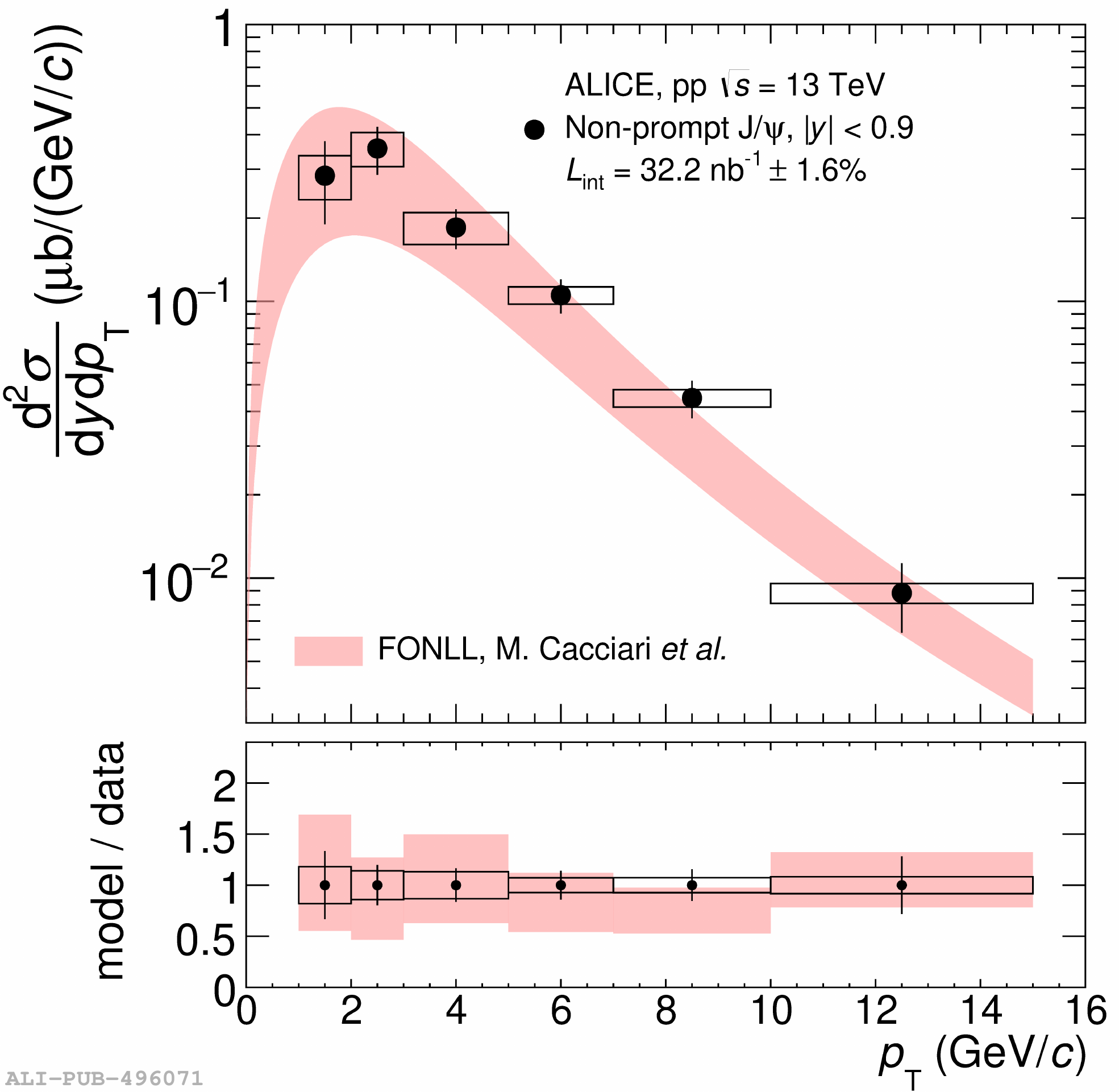} }
\end{center}
\caption{The prompt (left) and non-prompt (right) components of the $\mathrm{J} / \psi$ production cross section as a function of $p_{\mathrm{T}}$ at midrapidity and at $\sqrt{s}=13~\mathrm{TeV}$~\cite{PNP}, compared to model predictions~\cite{Cheung_2018, Ma_2014, Baranov_2019, Butensch_n_2011, Ma_2011, Cacciari_1998}.}
\label{fig:nonprompt}
\end{figure}

Figure~\ref{fig:nonprompt} shows the $p_{\mathrm{T}}$-differential cross section of the prompt and non-prompt $\mathrm{J} / \psi$ components, measured at midrapidity in pp collisions at $\sqrt{s}=13$ TeV~\cite{PNP}. The prompt contribution of the cross section (Fig.~\ref{fig:nonprompt}, left) is compared to expectations of ICEM~\cite{Cheung_2018}, NRQCD+CGC~\cite{Ma_2014}, NRQCD with $k_{\mathrm{T}}$ factorization~\cite{Baranov_2019}, and two other NLO NRQCD models~\cite{Butensch_n_2011, Ma_2011}.  The ICEM model shows good agreement with data at low $p_{\mathrm{T}}$ but slightly overshoots them at high $p_{\mathrm{T}}$. The Lipatov calculation based on $k_T$ factorization slightly overestimates data at low $p_{\mathrm{T}}$. The two NLO NRQCD calculations are compatible with data for $p_{\mathrm{T}} \geq 5$ GeV/$c$, while the NRQCD+CGC calculations show good agreement over the whole $p_{\mathrm{T}}$ range. The non-prompt component of the $\mathrm{J} / \psi $ cross section at $\sqrt{s}=13$ TeV is in good agreement with FONLL~\cite{Cacciari_1998} (Fig.~\ref{fig:nonprompt}, right).

\begin{figure}[h]
\begin{center}
\includegraphics[width=0.7\linewidth]{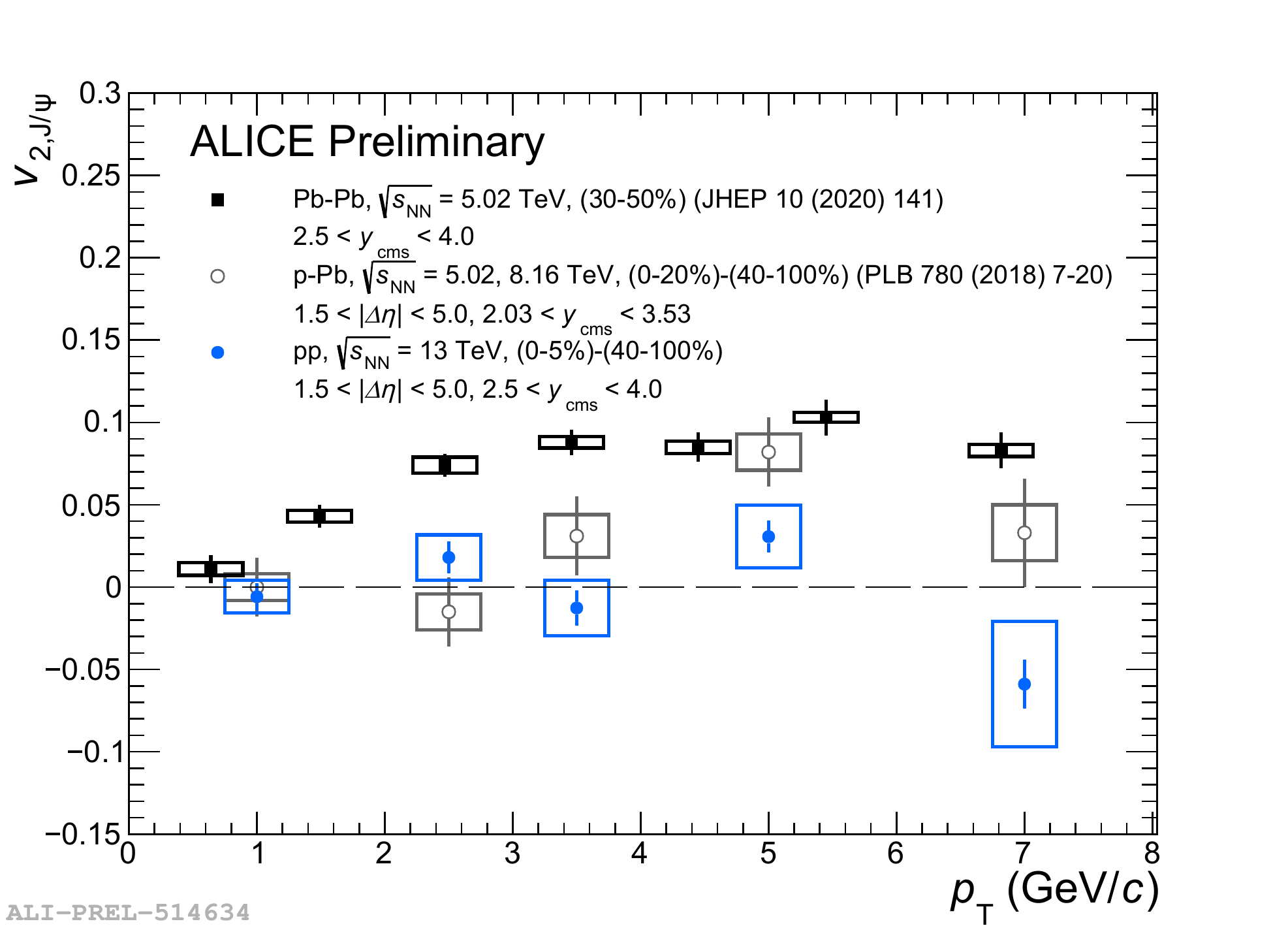} 
\end{center}
\caption{J/$\psi$ elliptic flow as a function of $p_{\mathrm{T}}$ in pp, p--Pb~\cite{Acharya_2018}, and Pb--Pb~\cite{2020} collision systems.}
\label{fig:v2}
\end{figure}

To investigate the participation of quarkonia in collective effects in small systems, the $\mathrm{J} / \psi$ elliptic flow ($v_{\mathrm{2}}$) was measured in high-multiplicity pp collisions at $\sqrt{s}=13$ TeV at forward rapidity in the dimuon decay channel. In heavy-ion collisions, the elliptic flow encodes the response of the hydrodynamic expansion of the produced QGP to the initial collision-geometry asymmetry, and quantifies the strength of collective phenomena in these collision systems. Similarly to the p--Pb case, the $v_{\mathrm{2}}$ measurement in pp collisions is obtained from the measurement of azimuthal correlations between inclusive $\mathrm{J} / \psi$ and charged hadrons. The result is presented in Fig.~\ref{fig:v2} and is compared to the p--Pb~\cite{Acharya_2018}, and Pb--Pb~\cite{2020} data. For these last two collision systems, the $\mathrm{J} / \psi$ shows a positive $v_{\mathrm{2}}$, indicating that it participates in the collective motion of the system. For p--Pb, the $\mathrm{J} / \psi$ elliptic flow shows similar values as in Pb--Pb for $p_{\mathrm{T}}$ > 4 GeV/$c$, which could hint at a common mechanism generating flow in the two collision systems. However, the TAMU transport model~\cite{Du_2015}, which describes $v_{\mathrm{2}}$ in Pb--Pb collisions, does not describe p--Pb data. Other calculations based on initial-state models such as the Color Glass Condensate (CGC) predict the presence of correlations among final-state particles that can give rise to flow-like effects~\cite{PhysRevD.102.034010}. In pp collisions the $\mathrm{J} / \psi$ $v_{\mathrm{2}}$ shows no significant deviation from zero as a function of $p_{\mathrm{T}}$. This conclusion remains valid when considering the result integrated over $p_{\mathrm{T}}$ in the range $ 1 \leq p_{\mathrm{T}} \leq 12 $ GeV/$c$, for which the deviation of $v_{\mathrm{2}}$ from zero is of the order of $1 \sigma$. Hence, no collective behavior is observed for the $\mathrm{J} / \psi$ in high-multiplicity pp collisions at the LHC, within current uncertainties.

\section{Summary}

The inclusive $\mathrm{J} / \psi$ $p_{\mathrm{T}}$-differential cross sections at forward rapidity in pp collisions at $\sqrt{s}=5.02$, 7, 8, and 13 TeV are described within uncertainties by with NRQCD+FONLL calculations within uncertainties. At midrapidity, the prompt $\mathrm{J} / \psi$ $p_{\mathrm{T}}$-differential cross section is described within uncertainties by various models such as NRQCD, NRQCD+CGC or ICEM, whereas FONLL describes well the non-prompt $\mathrm{J} / \psi$ data. In addition, the ICEM model describes within uncertainties the production cross sections of inclusive J/$\psi$ and $\psi$(2S) and of the three lower $\Upsilon$ states at forward rapidity, together with their energy dependence. Finally, the first measurement of the J/$\psi$ elliptic flow in high-multiplicity pp collisions at $\sqrt{s}=13$ TeV does not show evidence for the participation of J/$\psi$ in collective effects within current uncertainties.

\end{document}